\documentclass{iaus}
\usepackage{graphicx}

\title[Haro15 metallicity] 
{Haro15: Is it actually a low metallicity galaxy?}

\author[V. Firpo, G. Bosch, G. H{\"a}gele, A. I. D\'{\i}az \& N. Morrell]   
{Ver\'onica Firpo$^1$, Guillermo Bosch$^1$, Guillermo H{\"a}gele$^{1, 2}$,
\'Angeles I. D\'{\i}az$^2$, \and Nidia Morrell$^3$}

\affiliation{$^1$Facultad de Ciencias Astron\'omicas y Geof\'{\i}sicas,
  Universidad Nacional de La Plata,\\Paseo del Bosque s/n, B1900FWA,
  La Plata, Argentina.email: {\tt vfirpo@fcaglp.unlp.edu.ar} \\[\affilskip]
$^2$Departamento de F\'{\i}sica Te\'orica,C-XI, Universidad Aut\'onoma de Madrid, Spain.\\[\affilskip]
$^3$Las Campanas Observatory, Carnegie Observatories, La Serena, Chile.}

\pubyear{2009}
\volume{265}  
\pagerange{119--126}
\jname{Chemical Abundances in the Universe: Connecting First Stars to Planets}
\editors{K. Cunha, M. Spite \& B. Barbuy, eds.}
\begin{document}

\maketitle

\begin{abstract}
We present a detailed study of the physical properties of the nebular material
in multiple knots of the blue compact dwarf galaxy Haro 15. Using long slit
and echelle spectroscopy, obtained at Las
Campanas Observatory, we study the physical conditions (electron density and
temperature), ionic and total chemical abundances of several atoms, reddening
and ionization structure. The latter was derived by comparing the oxygen and
sulphur ionic ratios to their corresponding observed emission line ratios (the
$\eta$ and $\eta$' plots) in different regions of the galaxy. Applying
direct and empirical methods for abundance determination, we perform a
comparative analysis between these regions.
\end{abstract}

{\underline{\it Observations}}

\begin{figure*}[ht!]
\begin{center}
 \includegraphics[width=1.7in]{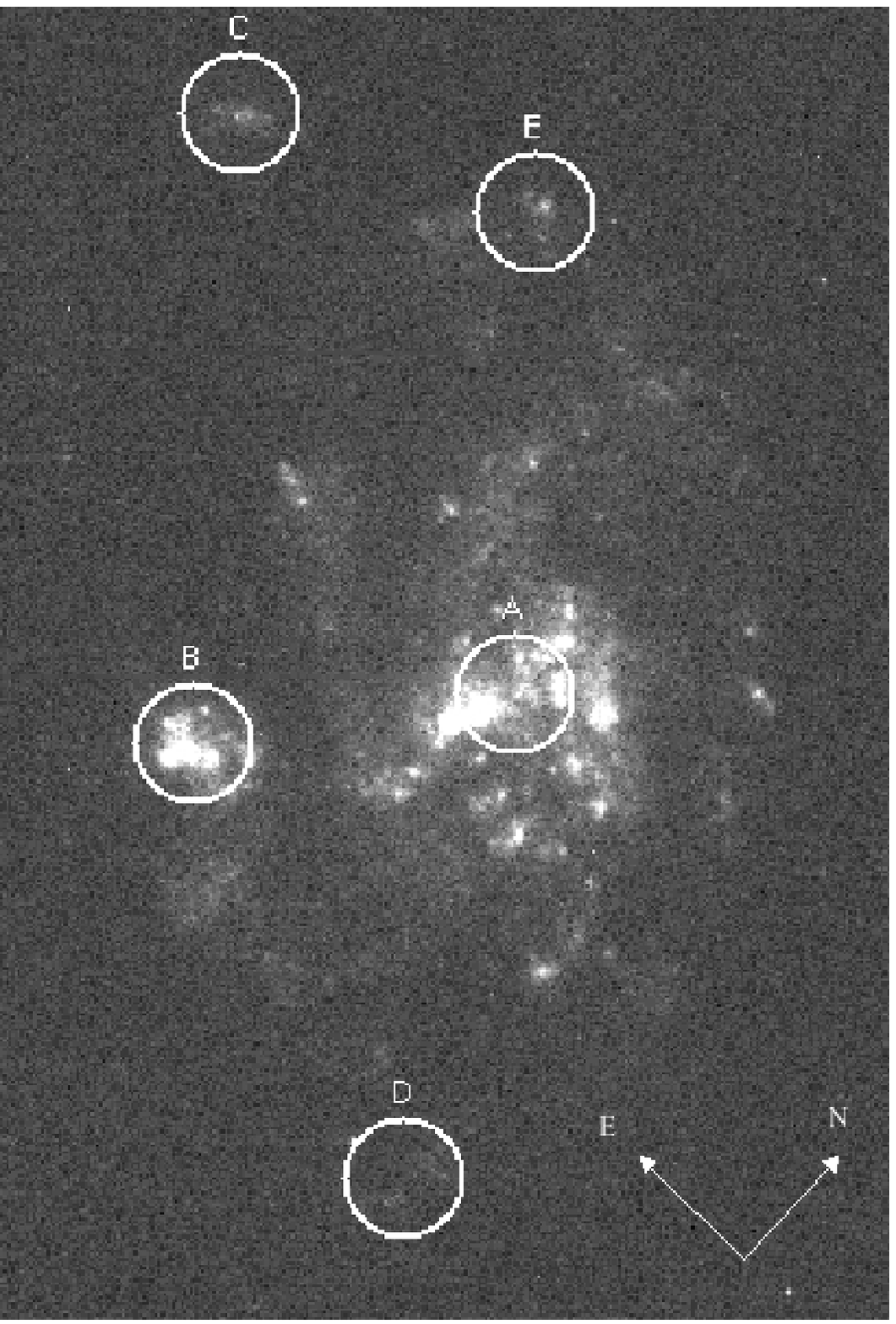} 
 \includegraphics[width=2.6in]{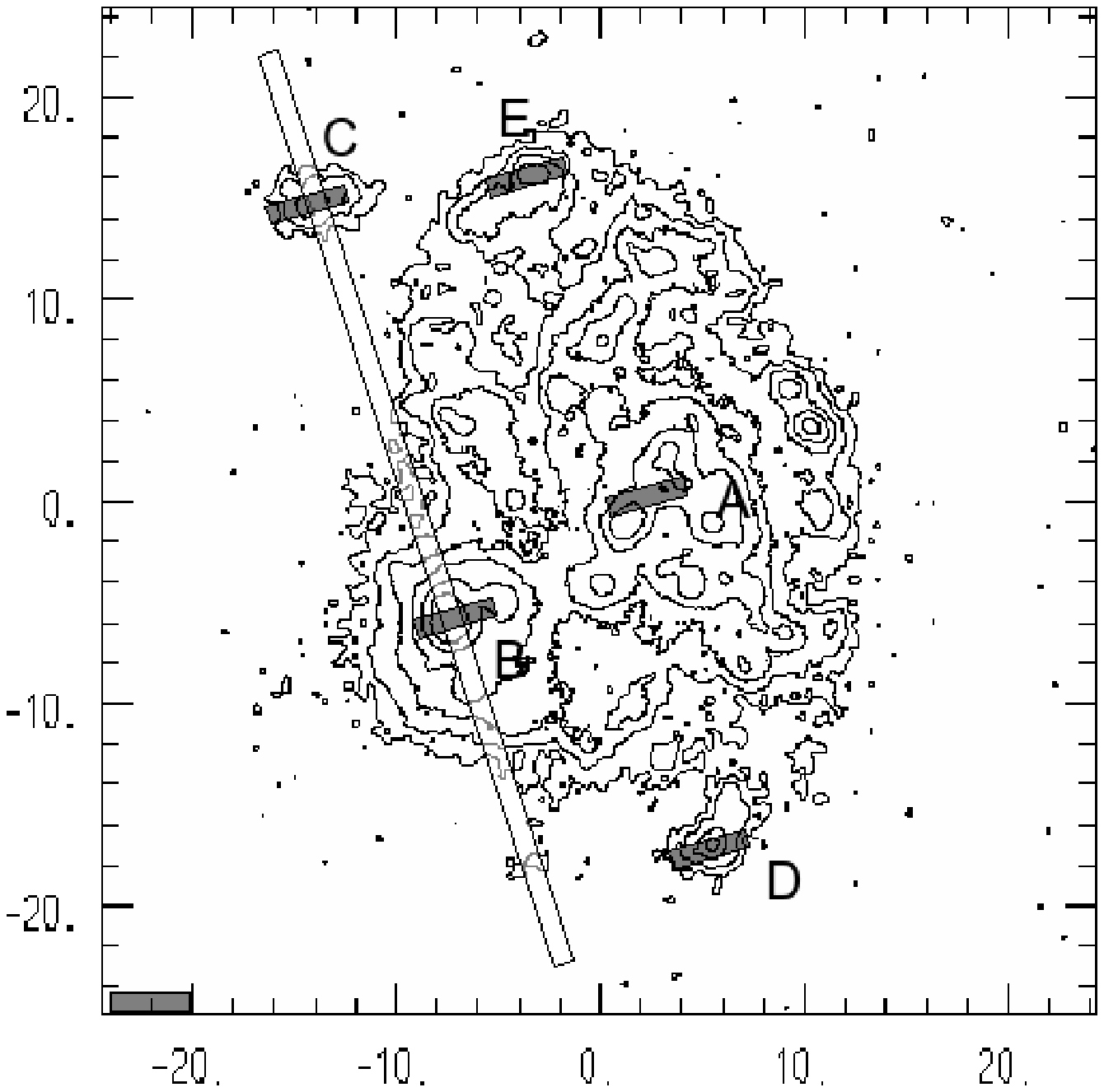} 
 \caption{Left panel: HST archive image showing the location of the observed regions.
Right panel: Contour maps of the continuum-subtracted H$\alpha$ image
(\cite[Cair{\'o}s et al. 2001]{cairos01}). In this image we show the slits
distribution. The  long-slit spectroscopy (1" wide) can be seen running
through Haro15-B and Haro15-C. Five pointings were observed with the echelle
mode, shown as dark rectangular (1"x4") slits.} 
   \label{fig}
\end{center}
\end{figure*}

We obtained high dispersion (Echelle Spectrograph; $\Delta\lambda$= 0.148\AA\ px$^{-1}$ at 5400\AA\, equivalent to 
6.72 kms$^{-1}$px$^{-1}$) and long-slit low resolution spectra (WFCCD; $\Delta\lambda$= 4.2\AA\ px$^{-1}$ at 5400\AA) of five 
luminous knots in the BCD galaxy Haro\,15 ( Fig.\,\ref{fig}) at the du Pont Telescope. We have
obtained a good flux calibration in both groups of data and we can confirm
that the two different data groups are comparable in knot B and C. 

{\underline{\it Electron density and temperatures}}

We have followed the same procedure described in detail in \cite[H\"agele et
al.\ (2006), (2008), H\"agele (2008)]{hagele06,hagele08,Hagele08_th} to derive the
physical properties of the nebular material in the observed regions. The
[OIII]\,$\lambda$\,4363\AA\ auroral emission line is only detected in knot 
B (longslit and echelle) and knot C (longslit). It was only possible to derive
T[OIII], T[SIII], T[OII] and T[SII] from direct 
measurements for knot B. For the other knots, for which we are not able to
measure some lines such as [OII]\,$\lambda\lambda$\,3727\AA,
[SIII]\,$\lambda\lambda$\,9069,9532\AA\ or auroral lines, we have resorted to
models that predict relationships between emission lines for different temperatures, for
example the relation between T[OII] and T[OIII] found by
\cite[P{\'e}rez-Montero \& D{\'{i}}az (2003)]{PMD03}. In the case of
densities, the knots show density values in the low density limit (ne $<$
100cm$^{-3}$), typical for of this kind of objects, with the exception of knot E (ne
$\approx$ 280cm$^{-3}$).

{\underline{\it Chemical Abundances: O, S, N, Ar, Ne}}

Oxygen abundances and their uncertainties were derived for each observed
knot using the direct method, where it could be applied, or several empirical
methods using the strong emission lines present in the spectra. We notice a
difference in the O/H ratio between knots A and B. This difference was
suggested by \cite[L{\'o}pez-S{\'a}nchez \& Esteban (2009)]{LS09} as
the two objects might have had a different chemical evolution. Our results
support these differences
between the two regions. Knot C shows oxygen abundance similar to that of knot
B, while the oxygen abundance derived for knot E is closer to the abundance
calculated for knot A. The S/N in our knot D spectra is not as good as for the
other knots, and
the quantities derived for this region should be used with caution.  

In the cases that we can use the direct method, the total abundances have been
derived taking into account the unseen ionization stages of each element,
resorting to the most widely accepted ionization correction factors (ICF) for
each species [X/H=ICF(X$^{+i}$) * (X$^{+i}$/H$^{+}$)] 
(see \cite[P{\'e}rez-Montero et al.\ 2007, H\"agele et
al.\ 2008]{PMD07,hagele08}). 
The N and Ar abundances are higher in knot A
than in knot B, although the S abundance derived for knot A lies between those
derived from echelle and longslit of knot B.  

{\underline{\it Ionization Structure}}

The ratio between O$^+$/O$^{2+}$ and S$^+$/S$^{2+}$ denoted by $\eta$ is
intrinsically related to the shape of the ionizing continuum and depends on
nebula geometry only slightly (\cite[V{\'{i}}lchez \& Pagel,
1988]{Vilchez88}). 
We can include knot B in this diagram, lying in the highest
excitation region. The purely observational counterpart, the 
$\eta$' diagram ($\eta$'=[([OII]/[OIII])/([SII]/[SIII])),
where $\eta$ and $\eta$' are related through the electron temperature but very
weakly. The position of knot B in both diagrams shows a compatible ionization
structure.

\end{document}